\documentstyle[12pt,times,epsf]{article}

\begin{document}

\baselineskip 6mm
\renewcommand{\thefootnote}{\fnsymbol{footnote}}

%------------ Sangmin's macro's, etc  -----------

\newcommand{\nc}{\newcommand}
\newcommand{\rnc}{\renewcommand}

%\headheight=0truein
%\headsep=0truein
%\topmargin=0truein
%\oddsidemargin=0truein
%\evensidemargin=0truein
%\textheight=9truein
%\textwidth=6.5truein

\rnc{\baselinestretch}{1.24}	% 1.5 spacing btwn text lines
\setlength{\jot}{6pt} 		% spacing btwn the rows of an eqnarray
\rnc{\arraystretch}{1.24}   	% spacing btwn the rows of a non-eqn array

%%%%%%%%%%%%%%%%%%%%%% Equation Numbering %%%%%%%%%%%%%%%%%%%%%%%
\makeatletter
\rnc{\theequation}{\thesection.\arabic{equation}}
\@addtoreset{equation}{section}
\makeatother                      

%%%%%%%%%%%%%%%%%%%%%%%%%%%%%%%%%%%%%%%%%%%%%%%%%%%%%%%%%%%%%%%%%
%								%
%		NEW COMMANDS AND MACROS				%
%								%
%%%%%%%%%%%%%%%%%%%%%%%%%%%%%%%%%%%%%%%%%%%%%%%%%%%%%%%%%%%%%%%%%

%%%%% Simplify some frequently used LaTeX commands %%%%%

\nc{\be}{\begin{equation}}
\nc{\ee}{\end{equation}}
\nc{\bea}{\begin{eqnarray}}
\nc{\eea}{\end{eqnarray}}
\nc{\xx}{\nonumber\\}

\nc{\eq}[1]{(\ref{#1})}
\nc{\newcaption}[1]{\centerline{\parbox{6in}{\caption{#1}}}}

\nc{\fig}[3]{
\begin{figure}
\centerline{\epsfxsize=#1\epsfbox{#2.eps}}
\newcaption{#3. \label{#2}}
\end{figure}
}

%%%%% Journal Macros %%%%%

\nc{\np}[3]{Nucl. Phys. {\bf B#1} (#2) #3}
\nc{\pl}[3]{Phys. Lett. {\bf #1B} (#2) #3}
\nc{\prl}[3]{Phys. Rev. Lett.{\bf #1} (#2) #3}
\nc{\prd}[3]{Phys. Rev. {\bf D#1} (#2) #3}
\nc{\ap}[3]{Ann. Phys. {\bf #1} (#2) #3}
\nc{\prep}[3]{Phys. Rep. {\bf #1} (#2) #3}
\nc{\rmp}[3]{Rev. Mod. Phys. {\bf #1} (#2) #3}
\nc{\cmp}[3]{Comm. Math. Phys. {\bf #1} (#2) #3}
\nc{\mpl}[3]{Mod. Phys. Lett. {\bf #1} (#2) #3}
\nc{\cqg}[3]{Class. Quant. Grav. {\bf #1} (#2) #3}
\nc{\jhep}[3]{J. High Energy Phys. {\bf #1} (#2) #3}

%%%%% Special Letters

\def\vare{\varepsilon}
\def\bz{\bar{z}}
\def\bw{\bar{w}}

%%% Caligraphic letters %%%%

\def\CA{{\cal A}}
\def\CC{{\cal C}}
\def\CD{{\cal D}}
\def\CE{{\cal E}}
\def\CF{{\cal F}}
\def\CG{{\cal G}}
\def\CT{{\cal T}}
\def\CM{{\cal M}}
\def\CN{{\cal N}}
\def\CP{{\cal P}}
\def\CL{{\cal L}}
\def\CV{{\cal V}}
\def\CS{{\cal S}}
\def\CW{{\cal W}}
\def\CY{{\cal Y}}
\def\CS{{\cal S}}
\def\CO{{\cal O}}
\def\CP{{\cal P}}
\def\CN{{\cal N}}

%%% Double line letters %%%

\def\IR{{\hbox{{\rm I}\kern-.2em\hbox{\rm R}}}}
\def\IB{{\hbox{{\rm I}\kern-.2em\hbox{\rm B}}}}
\def\IN{{\hbox{{\rm I}\kern-.2em\hbox{\rm N}}}}
\def\IC{\,\,{\hbox{{\rm I}\kern-.59em\hbox{\bf C}}}}
\def\IZ{{\hbox{{\rm Z}\kern-.4em\hbox{\rm Z}}}}
\def\IP{{\hbox{{\rm I}\kern-.2em\hbox{\rm P}}}}
\def\IH{{\hbox{{\rm I}\kern-.4em\hbox{\rm H}}}}
\def\ID{{\hbox{{\rm I}\kern-.2em\hbox{\rm D}}}}

%%% Greek letters %%%

\def\a{\alpha}
\def\b{\beta}
\def\ga{\gamma}
\def\d{\delta}
\def\ep{\epsilon}
\def\ph{\phi}
\def\k{\kappa}
\def\l{\lambda}
\def\m{\mu}
\def\n{\nu}
\def\th{\theta}
\def\rh{\rho}
\def\s{\sigma}
\def\t{\tau}
\def\w{\omega}
\def\G{\Gamma}

%%%%% Mathematical Symbols

\def\half{\frac{1}{2}}
\def\imp{\Longrightarrow}
\def\dint#1#2{\int\limits_{#1}^{#2}}
\def\goto{\rightarrow}
\def\para{\parallel}
\def\brac#1{\langle #1 \rangle}
\def\del{\nabla}
\def\grad{\nabla}
\def\curl{\nabla\times}
\def\div{\nabla\cdot}
\def\p{\partial}
\def\e{\epsilon_0}

%%%%% Roman font in math

\def\Tr{{\rm Tr}}
\def\det{{\rm det}}

%%%%% Names

\def\Kahler{K\"{a}hler}

%%%%% For this paper only

\def\e{\varepsilon}
\def\bA{\bar{A}}
\def\c{\zeta}

\begin{titlepage}
%---------------- preprint number ---------------
\hfill\parbox{4cm}
{hep-th/0003145 \\ KIAS-P00013 }

%------------------------ title ------------------------
\vspace{25mm}
\centerline{\Large \bf UV/IR Mixing in Noncommutative Field Theory}
\vskip 5mm
\centerline{\Large \bf via Open String Loops}
%---------------- authors and addresses ----------------
\vspace{10mm}
\begin{center}  
Youngjai Kiem and
Sangmin Lee\footnote{ykiem, sangmin@kias.re.kr} 
\\[10mm] 
{\sl School of Physics \\ Korea Institute for Advanced Study \\
Seoul 130-012 Korea}
\end{center}
\thispagestyle{empty}
\vskip 30mm

%----------------------- abstract ----------------------
\centerline{\bf ABSTRACT}
\vskip 5mm
\noindent
We explicitly evaluate one-loop (annulus) planar 
and nonplanar open string amplitudes in the presence
of the background NS-NS two-form field.  
In the decoupling limit of Seiberg and Witten,
we find that the nonplanar string amplitudes reproduce 
the UV/IR mixing of noncommutative field theories.  
In particular, the investigation of the UV regime of 
the open string amplitudes shows that certain IR closed 
string degrees of freedom survive the decoupling limit 
as previously predicted from the noncommutative 
field theory analysis.  These degrees of freedom are 
responsible for the quadratic, linear and logarithmic 
IR singularities when the D-branes embedded in 
space-time have the codimension zero, one and two, respectively.  
The analysis is given for both bosonic and supersymmetric open strings. 
\vspace{2cm}
\end{titlepage}

%-------------------------------------------------------
\baselineskip 7mm
\renewcommand{\thefootnote}{\arabic{footnote}}
\setcounter{footnote}{0}

\section{Introduction}

Certain noncommutative field theories \cite{connes}
can be systematically derived from open string 
theories in the presence of constant background 
NS-NS two-form field ($B$ field) \cite{shei}-\cite{witten}.  
The upshot of these developments is that noncommutative 
field theories are more stringy than what one might naively
expect.  For example, unlike the generic commutative field 
theories arising as decoupling limits of string theories, 
noncommutative field theories are T-duality 
invariant signaling its stringy nature \cite{doughull}.
Further considerations of loop effects in noncommutative
field theories \cite{filk,seiberg1,seiberg2} add an 
intriguing new element in the analogy between open 
string theories and noncommutative field theories, namely, 
the UV/IR mixing.  From the open
string perspective, the simplest one-loop annulus diagram
reveals a prominent stringy character, the
open/closed string channel duality.  The UV regime
of the open string annulus amplitudes can naturally be 
interpreted as the IR closed string degrees of freedom.
This behavior closely parallels the UV/IR mixing of 
Refs.~\cite{seiberg1,seiberg2} in noncommutative field
theories, where one can interpret
certain UV divergences coming from nonplanar loops
of high energy virtual particles as IR divergences.   

In this paper, we embark upon the detailed study of
one-loop annulus open string amplitudes in the presence
of constant background $B$ field\footnote{We note
that our calculations have overlaps with the earlier
literature on open string amplitudes, such as
Refs.~\cite{fradtsey} and \cite{clny}.} and 
recover many features found in 
Refs.~\cite{seiberg1,seiberg2}.  The one-loop open
string amplitudes turn out to be of the same form
as Ref.~\cite{seiberg1}, except including the 
contributions from the massive string excitations.
Upon taking the decoupling limit of \cite{witten},
massive excitations decouple,
while some UV degrees of freedom do not
in nonplanar diagrams.  Since our set-up is the 
string theory
framework, via the standard open/closed string 
duality, we can unambiguously identify these
extra degrees of freedom as IR 
closed string contributions.
Their Wilsonian effective action decoded 
from the
annulus amplitudes also turns out to be the same
as the one proposed in Ref.~\cite{seiberg1} for
the extra degrees of freedom responsible for 
the IR singularities of the one-loop noncommutative field 
theory amplitudes.  In particular, for $(D-1)$
D-branes (original critical open string theory),
$(D-2)$-branes and $(D-3)$-branes, open string 
theory calculations reproduce the 
quadratic divergences \cite{seiberg1}, linear
divergences and logarithmic divergences \cite{seiberg2}
caused by the extra degrees of freedom, respectively, where
$D$ is the dimension of space-time.  In short, 
noncommutative quantum field theories arising as 
limits of open string theories include closed string
degrees of freedom, which survive the decoupling
limit, couple linearly to the D-brane world-volume open
string degrees of freedom and live in the bulk space-time.

This paper is organized as follows.  In section 2, 
we compute the world-volume propagators on an annulus
in the presence of the constant background $B$
field.  In section 3, we evaluate the planar
and nonplanar annulus diagrams in the bosonic
open string theory.
Via open/closed string duality, we 
identify the IR closed string degrees of freedom,
which survive the decoupling limit, and study their
properties.  In section 4, we extend our analysis 
to open superstring amplitudes.  We discuss  
further directions and implications suggested by
our analysis in section 5.

%%%%%%%%%%%%%%%%%%%%%%%%%%%%%%%%%%%%%%%%%%%%%%%%%%
%
\section{World-sheet propagator on an annulus}
%
%%%%%%%%%%%%%%%%%%%%%%%%%%%%%%%%%%%%%%%%%%%%%%%%%%

One important ingredient in computing the one-loop string amplitude 
is the world-sheet propagator on an annulus. 
It was first obtained in \cite{clny} using a world sheet coordinate 
in which the two boundaries of the annulus are concentric circles. 
Here we present an equivalent but more concise form of the propagator 
following the notations of \cite{rey}.

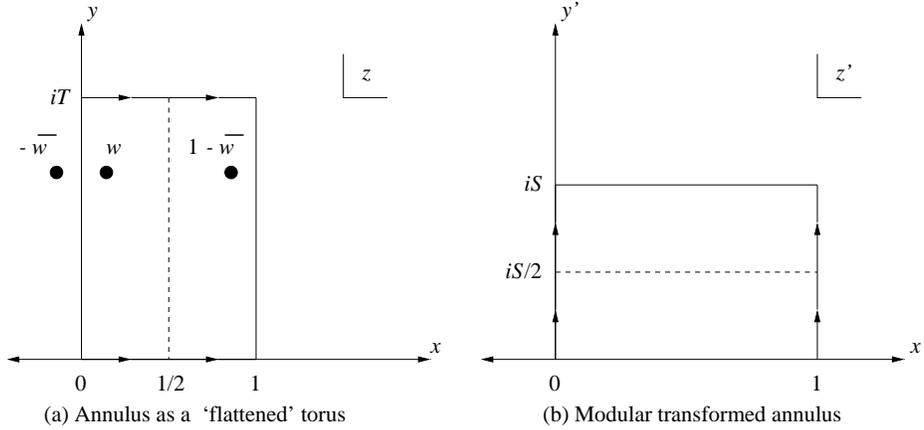
\fig{350pt}{annulus}{The world sheet coordinate for annulus}

First, in the absence of the $B$ field, 
consider a rectangular torus whose modulus parameter $\tau=iT$ 
is purely imaginary. The world-sheet propagator is 
\begin{equation}
 \langle X^\m(z) X^\n(w) \rangle = \frac{\a'}{2} \eta^{\m\n}G (z-w),
\end{equation}
where 
\begin{equation}
 G( \nu ) = - \log \left| \frac{ \theta_1 (\nu | iT ) }
  { \theta_1^{\prime} ( 0 | iT ) } \right|^2 + \frac{2 \pi}{T}
  \left[ {\rm Im } ( \nu ) \right]^2 .
\end{equation}
Here, $\theta_1$ is the theta function defined as
\[ \theta_1 ( \nu | \tau ) = -i \sum_{m = -\infty}^{\infty}
 (-1)^m q^{\frac{1}{2} ( m + \frac{1}{2} )^2 } \omega^{m + 
\frac{1}{2} } ~, \]
\[ q = \exp ( 2 \pi i \tau ) ~~~ , ~~~ \omega = 
       \exp ( 2 \pi i \nu ) ~, \]
The propagators are periodic
under the two lattice transformations
\[
z \goto z+1,\  \  z \goto z+iT 
\]
and they satisfy the flux conversation; the integral of 
$\partial^2 G (z)$ over the torus vanishes.
To turn torus propagators
to annulus propagators, we place a mirror charge
at $- \bar{w}$ (and at all their lattice translation points)
for a source charge at $w$.  This operation imposes 
Neumann boundary conditions along the two boundaries
at ${\rm Re} (z) = 0$ and ${\rm Re} (z) = 1/2$, while 
maintaining the periodicity in $z \goto z+iT$, 
thereby turning the original torus to an annulus.  
Written explicitly, the propagators look like
\begin{equation}
 \langle X^\m (z) X^\n (w) \rangle = 
 \frac{\a'}{2} \eta^{\m\n}\{G (z-w) + G (z+\bar{w}) \}.
\label{ann}
\end{equation}
The propagators with Dirichlet boundary condition
can also be straightforwardly written down.  However,
we will not need them for we will consider only
open string vertex insertions.

Our next task is to find the explicit form of analogous 
expressions that are valid when $B \ne 0$. 
As noted in \cite{hkll}, one can bring the $B$ 
field into a block-diagonal 
form and consider each $2\times 2$ block separately. 
Suppose for now that we turn on the $B_{12} = B $ 
along $X^1 = X$ and $X^2 = Y$ directions parallel to the D-branes 
under consideration. 
The boundary conditions at ${\rm Re} (z) = 0$ and ${\rm Re} (z) = 1/2$
should be modified into
\footnote{
As in \cite{clny}, it is possible to trade the quadratic
terms of (\ref{xxpro}) with modified
boundary conditions involving a constant term on the 
right hand sides of (\ref{modbc}).  
Our choice in this paper is
to keep the boundary conditions (\ref{modbc}) intact.
} 
\begin{equation}
 \partial_n X + i B \partial_t  Y = 0 |_{{\rm Re} (z) = 0, 1/2}.
\label{modbc}
\end{equation}
The answer is:
\bea
\frac{2}{\a'} \langle X(z) X(w) \rangle 
 &=& G(z-w) + \frac{1-B^2}{1+ B^2} G(z+ \bar{w}) 
+ \frac{ B^2}{1+ B^2} \frac{4 \pi}{T} 
 \left[ {\rm Re } ( z+ \bar{w} ) \right]^2 ~,
\label{xxpro} 
\\
\frac{2}{\a'} \langle X(z) Y(w) \rangle 
 &=& \frac{2B}{1+ B^2} \left[ \log \frac{ \theta_1 ( z+ \bar{w} ) }
 { \theta_1 (\bar{z} + w ) } + \frac{4 \pi i}{T} 
  {\rm Re} (z+ \bar{w} ) {\rm Im } ( z+ \bar{w} ) 
  \right] ~. 
\label{xypro}
\eea
When $|z|,|w| \ll 1, T$, the quadratic terms are negligible and 
$\theta_1 (z | iT)$ reduces to $z$, so that we recover the propagators 
on a disk, which was obtained, for example, in
Refs.~\cite{schom,witten,hkll}.  
The coefficients of the $\theta$-function terms are uniquely 
determined by comparison with the propagator on the disk. 
The quadratic terms are required by the
periodicity in $z \goto z+ iT$ 
and flux conservation\footnote{The periodicity in this context
means the periodicity of the physical objects, such as
$ \langle \partial X Y \rangle$.  We note that 
$ \langle X Y \rangle $ itself is not periodic, but this
does not give an ambiguity when computing physical
amplitudes.}.

For the computation of the open string amplitudes, we 
insert open string vertex operators along 
the boundaries at ${\rm Re} (z) = 0$ and ${\rm Re} (z) = 1/2$. 
For later convenience, we note that for the planar
insertions, the propagators become
\bea
\langle X(0) X(iy) \rangle
&=& \a' \frac{1}{1+B^2} G(iy) ~, \label{puha}
\\
\langle X(0) Y(iy) \rangle
&=&
i ( \pi \a' ) \frac{ B}{1+B^2} \e(y) ~,
\eea
where we introduced the Heaviside step function 
$\epsilon ( y)  = y / | y |$. 
For nonplanar insertions, we have
\bea
\langle X(0) X(1/2 + iy) \rangle
&=&
\a' \frac{1}{1+B^2} G(1/2 + iy) 
 + \frac{\pi \a'}{2T} \frac{B^2}{1+B^2} ~, 
\\
\langle X(0) Y(1/2 + iy) \rangle
&=& 
i (2 \pi \a') \frac{ B}{1+B^2} \frac{y}{T} ~. \label{puhaha}
\eea
The quadratic term in \eq{xxpro}, 
\begin{equation}
 \frac{2 B^2}{1+ B^2} \frac{2 \pi}{T} 
 \left[ {\rm Re } ( z+ \bar{w} ) \right]^2 ~ ,
\label{winding}
\end{equation}
deserves further comments.
We note that (\ref{winding}) distinguishes between the planar and
nonplanar vertex insertions.  In particular, it vanishes
for planar insertions along ${\rm Re} (z) = 0$ 
but gives a non-vanishing contribution
for nonplanar insertions where one vertex is separated
from another.  We will find that it will give 
finite contributions to the amplitudes
under the decoupling limit of Ref.~\cite{witten}
thus surviving in the effective noncommutative field theory.

%%%%%%%%%%%%%%%%%%%%%%%%%%%%%%%%%%%%%%%%%%%%%%%%%%%%%%%%%%%%%
%
\section{One-loop bosonic open string amplitudes}
%
%%%%%%%%%%%%%%%%%%%%%%%%%%%%%%%%%%%%%%%%%%%%%%%%%%%%%%%%%%%%%

Utilizing the world-sheet propagators of section 2, we
explicitly evaluate the annulus amplitudes, inserting
two open string vertex operators along the world-sheet 
boundaries in the presence of D-branes.
In the following, we will ignore all numerical constants
in the overall normalization of the amplitudes, 
but the dimensions and dependence on coupling constant 
will be unambiguous. 

\subsection{Planar and nonplanar bosonic open string 
amplitudes on an annulus}

The one-loop amplitude in the 
presence of an $(n-1)$-brane is given by
\bea
{\cal A} 
&=& 
\int_0^\infty \frac{dT}{T} Z(T) 
\int_0^T dy_1 \int_0^T dy_2
\langle V_1(p,y_1) V_2(-p,y_2) \rangle_T \\
&=& 
g^2 \a' \int_0^{\infty} \frac{dT}{T} (2\pi\a' T)^{-n/2}
  f_1 (q)^{-24} T \int_0^T dy \;\; I(p;y,T) ~.
\label{genampl}
\eea 
The ``partition function'' part is computed in the same way 
as in \cite{pol};
\bea
Z(T) &=& \int \frac{d^n k}{(2\pi)^n} 
\sum_I e^{-2\pi\a' T(k^2+M_I^2)}, \\ 
f_1(q) &=& q^{1/24} \prod_{m=1}^\infty (1-q^m),\  \ q = e^{-2\pi T}.
\eea
The vertex operators for a tachyon and a gauge boson are, respectively,
\be
V_T = g\sqrt{\a'} e^{ikX}, V_A = g \e\cdot\partial X e^{ikX} ~,
\ee
where the coupling constant $g$ is the one that appears in the 
low energy effective action of open strings. Schematically,
\be
\label{leea}
S = -\int d^n x \{ (\partial \phi)^2 +m^2\phi^2 + g^2 \phi^4 + \cdots \} 
\ee
A simple dimensional analysis shows that $g$ is related to string coupling by 
$g^2 = ( \a' )^{n/2 -2} g_{st}$.

We first give the answers for the planar amplitudes.
When $B=0$, for the tachyon-tachyon insertion, we have
\begin{equation}
I = \left| \frac{\theta_1 (iy | iT) }{\theta_1^{\prime}
 (0 | iT ) } \right|^{-2} \exp \left( \frac{2 \pi }{T} y^2  \right) ~,
\label{tachb0}
\end{equation}
while for gauge boson-gauge boson insertion, we have
\be
\begin{array}{rcl}
I &=& 
\{ \epsilon_1 \cdot \epsilon_2 J'' 
+ \a' (\e_1\cdot p) (\e_2\cdot p) (J')^2 \}  e^{\a' p^2 J} \\
&=&
- \a' \{ (\e_1 \cdot \e_2) p^2 - (\e_1\cdot p) (\e_2\cdot p) \} (J')^2
e^{\a' p^2 J} + \mbox{(total derivative in $y$)} ~,
\end{array}
\label{pgaugeb}
\ee
where we define $J(y) = G(iy)$.
When we turn on the $B$ field ($B \ne 0$), the answer
is exactly the same as the one for $B=0$, except that the external
momentum squared is evaluated with respect to the open
string metric $G^{\mu \nu}$ defined in Ref.~\cite{witten}
as
\begin{equation}
G_{\m\n} \equiv \eta_{\m\n} - (B\eta^{-1}B)_{\m\n} ~ .
\end{equation}

We now consider the nonplanar insertions where $B \ne 0$
effect is conspicuous.  For the
tachyon-tachyon insertions, we get
\begin{equation}
I =   \left| \frac{\theta_2 ( iy | iT) }
{\theta_1^{\prime}
 (0 | iT ) } \right|^{-2} 
\exp \left( \frac{2 \pi }{T} y^2 \right) 
\exp \left( - \frac{p \circ p}{2\pi\a' T} \right)  ~,
\label{nptach}
\end{equation}
and the gauge boson-gauge boson insertions yield
\be
\begin{array}{rcl}
I &=& 
- \left[
\a' \{ (\e_1 \cdot \e_2) p^2 - (\e_1\cdot p) (\e_2\cdot p) \} (K')^2
+ \frac{(\e_1\times p)(\e_2\times p)}{\a'T^2}
\right]
\\
&& 
\times
\exp\left( \a' p^2 K - \frac{p \circ p}{2\pi\a' T} \right) ~,
\end{array}
\label{npgaugeb}
\ee
where we define $K(y) = G( 1/2 + iy )$.
We introduced $\circ$-product and $\times$-product
following Ref.~\cite{seiberg1} as
\be
 p \circ p \equiv - \frac{1}{4} p_\m (\Theta G \Theta)^{\m\n} p_\n
 ~~~ , ~~~
 \epsilon \times p = \epsilon_{\mu } \Theta^{\mu \nu}  p_{\nu} ~,
\ee
where $\Theta$ is
the noncommutativity parameter defined in \cite{witten},
\be
\Theta^{\m\n} \equiv -2\pi\a' \{(\eta+B)^{-1}B(\eta-B)^{-1}\}^{\m\n}.
\ee
The sign in the definition of $p\circ p$ is introduced
to make it nonnegative.

The $B$ dependence in the amplitudes come from two
combinations $p \circ p$ and $ \epsilon \times p$.
Due to the prefactor $1/T$ in front of $p \circ p$, 
the effect of noncommutativity becomes stronger 
as we approach the UV corner of the moduli integral.
For the higher spin world-volume fields, there are
polarization dependences, as exemplified in
$\epsilon \times p$ for the gauge boson amplitudes.

\subsection{Open/Closed string duality and the Decoupling limit}

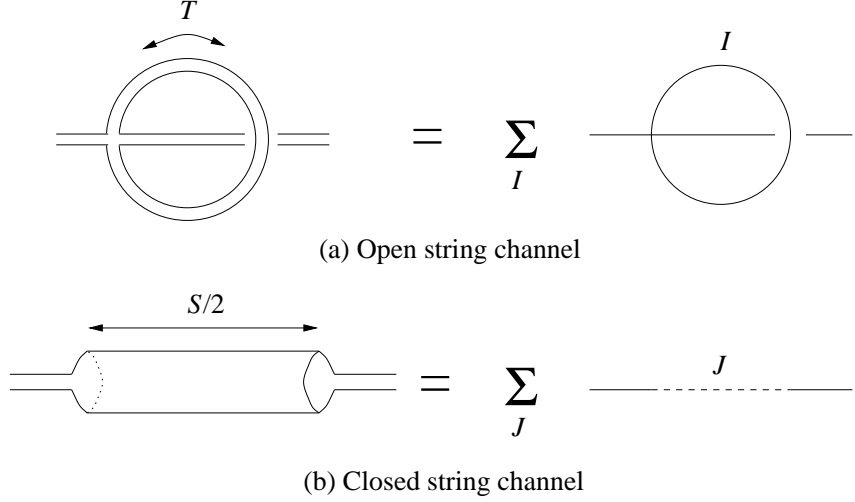
\fig{320pt}{channel}{Open/Closed string channel duality}

For the rest of this section, we only consider the 
tachyon amplitudes in detail for simplicity.
We first review the well-known 
world-sheet duality for $B=0$ to contrast 
it with the $B \ne 0$ situation.
The relation between the nonplanar 
tachyon-tachyon amplitude 
\be
\label{btr}
\CA = \int_0^\infty \frac{dT}{T} (2\pi\a'T)^{-n/2} T
\int_0^T dy 
\left| \frac{\theta_2 (iy | iT) }{\theta_1^{\prime}
 (0 | iT ) } \right|^{-2} 
\exp \left( \frac{2 \pi }{T} y^2\right)
\ee
and the one-loop amplitudes in the field theory of the type \eq{leea} 
is manifest in the region $T \gg 1$, where we can expand \eq{btr} in 
$e^{-2\pi T}$. For example, when $y \ll T$, the open string diagram 
looks very much like the field theory diagram in 
Fig.~\ref{channel}(a). 
This intuitive picture is confirmed by an explicit calculation 
which shows that
\be
\CA = g^2\a' \sum_I a_I
\int\frac{dT}{T} (2\pi \a'T)^{-n/2} T  e^{-2\pi\a' T M_I^2}
= \sum_I \int d^n k \frac{a_I g^2}{k^2 +M_I^2},
\ee
where $a_I$ are some numerical coefficients

In the opposite end $T \ll 1$, 
the usual channel duality allows us to rewrite \eq{btr}
from the point of view of the closed strings.  
In particular, using the modular transformation of the theta functions, 
we find 
\be
\label{opp}
{\cal A} = \int_0^{\infty} \frac{dS}{S} S^{n/2 -12} 
 f_1 ( \tilde{q} )^{-18}
\int_0^1 dx \left| \theta_4 (x | iS ) \right|^{-2}  ~ ,
\label{cc}
\ee
where $S = 1/T$ and $\tilde{q} = \exp ( - 2\pi S )$. 
The picture now is a closed string connecting the open string 
states as in Fig.~\ref{channel}(b).
For the case of the space-time filling 25-brane, 
\eq{opp} can be expanded to give
\be
\CA = 
\sum_J b_J \int dS e^{-2\pi S \a'(p^2 +M_J^2)/4} \sim
\sum_J \frac{b_J^2 \kappa^2}{p^2 +M_J^2}.
\label{temp1}
\ee
for some numerical constants $b_J$.
\footnote{
When computing perturbative string amplitudes, the external 
momenta are always put on-shell.
Here we are assuming that the final expression holds for off-shell 
amplitudes as well.
}
The coupling constant $\kappa$ appears in the low energy effective action 
of the form
\be
S = \int d^n x \kappa \chi\phi,
\ee
where $\chi$ is a closed string field.

The noncommutative field theory arises 
in the decoupling limit $\a' \goto 0$ while 
keeping $G^{\m\n}$ and $\Theta^{\m\n}$ fixed \cite{witten}.
We note that in the bosonic string theory, the mass spectrum 
is known to be
$\a'M_I^2 = N_I-1$ (open) and $\a'M_J^2 = 4(N_J-1)$ (closed)
for nonnegative integers $N_I$ and $N_J$. 
In this limit, therefore, if we ignore the tachyons, 
all but the contribution from massless intermediate 
states disappear, as can be seen from (\ref{temp1}).
When $B=0$, the massless intermediate degrees of freedom
give a trivial IR divergence that should be
cancelled with other divergences via Fischler-Susskind
type mechanism.  When $B \ne 0$, their contribution
is non-trivial as we will see shortly; we need to 
take the Wilsonian point of view regarding the cutoff
and the effective degrees of freedom.   

\subsection{What survives the decoupling limit}

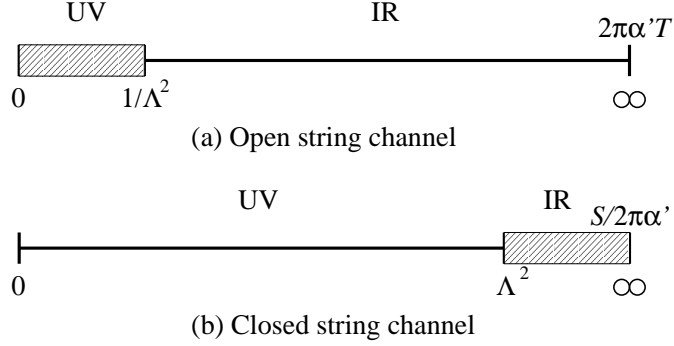
\fig{250pt}{cutoff}{Domain of moduli integral}

The key issue is to identify the contributions from
the $T \ll 1 $ UV regime to the open string moduli 
integral when $B \ne 0$.  
For this purpose, in the spirit of 
string field theory \cite{witten2,sft1}, 
we explicitly introduce a short distance UV regulator $1/\Lambda^2$ 
in the open string description; the regulated open string 
contribution comes from the region of the moduli space where 
$2 \pi \alpha^{\prime} T > 1/\Lambda^2$.
Then, as depicted in Fig.~\ref{cutoff}, the possible 
extra UV degrees of freedom
originating from the extreme UV open string loops should come 
from the corner of moduli space where 
$ 0 < 2 \pi \alpha^{\prime} T < 1/\Lambda^2$.  
When $2 \pi \alpha^{\prime} T$ goes below the UV cutoff, 
we have a factorization channel where the original 
nonplanar annulus diagram becomes two string states 
connected by a long  closed string tube
\footnote{
In \cite{sft2}, 
the same amplitude was computed using the 
open-closed string field theory for $B=0$. 
Division of the moduli space into two connected parts
is inherent in their formalism.
}.  
Via open/closed string channel duality, it is natural to
investigate  $ 0 <2 \pi \alpha^{\prime} T < 1/\Lambda^2$ 
corner of the open string
moduli space in terms of the closed string picture.  We thus
resort to the nonplanar amplitude expression
in the closed string channel, Eq.~(\ref{cc}).
In this channel, the open string UV cutoff $1/\Lambda^2$ 
transforms to the closed string IR cutoff $\Lambda^2$.
As shown in Fig.~\ref{cutoff}, the open string UV regime 
gets mapped to the closed string IR regime 
$ S/ 2 \pi \alpha^{\prime} > \Lambda^2 $.  The contribution
to the nonplanar amplitude from these IR closed string degrees
of freedom can be computed as
\begin{equation}
{\cal A}_{\rm IR} = {\cal A } ( \infty ) - {\cal A} ( \Lambda )
 \simeq \int_0^{\infty} \frac{dS}{S} S^{n/2 - 12} 
( e^{ - p \circ p S / 2 \pi \alpha^{\prime} }  - 
 e^{ - ( p \circ p + 1/ \Lambda^2 ) S / 
2 \pi \alpha^{\prime}} )  ~,
\label{edeg}
\end{equation}
where the IR regulated closed string amplitude ${\cal A}
 ( \Lambda )$ is defined as
\begin{equation}
{\cal A } ( \Lambda )
 \simeq \int_0^{\infty} \frac{dS}{S} S^{n/2 - 12} 
e^{ - ( p \circ p  
  +  1 / \Lambda^2  ) S / 2 \pi \alpha^{\prime} } ~, 
\label{careg}
\end{equation}
explicitly introducing the cut off $\Lambda^2$.  The
${\cal A} (\Lambda )$ defined in (\ref{careg}) is the IR 
regulated amplitude where we restrict the closed string moduli 
integral to the distance up to the IR cutoff scale $\Lambda^2$.  
In (\ref{edeg}), we wrote down only the part of the amplitude
that {\em survives the decoupling limit} and we deleted the 
tachyonic intermediate contribution.  
Reinstating the tachyonic contribution 
would produce 
the negative eigenvalue for the quadratic effective
action for the small value of $p \circ p$
by shifting 
$ p \circ p$ into $p \circ p + M_{tachyon}^2 $
in (\ref{edeg}),
indicating the tachyonic instability \cite{seiberg1}.     
For the space-time filling 25-brane ($n=26$), we have
\begin{equation}
{\cal A}_{\rm IR} = 
 \kappa^2 (\a')^2 
 \left[ \frac{1}{ p \circ p} - \frac{1}{ p \circ p + 1/ \Lambda^2 } \right]
 = \frac{\kappa^2(\a')^2}{ p \circ p + \Lambda^2 ( p \circ p )^2 } ~ 
\label{eff1}
\end{equation}
from (\ref{edeg}).  From the `long tube' IR closed string
picture of Fig.~2, we find that (\ref{eff1}) is nothing
but the propagator of the extra degree of freedom multiplied
by the coupling constant.  From the low energy effective
description point of view, the extra degree of freedom
(denoted as $\chi$ field) then has the effective Lagrangian 
of the form
\begin{equation}
 \int dx^{26} \left[ \partial \chi  \circ  \partial \chi
 + \Lambda^2 ( \partial \circ \partial \chi )^2 \right]
 + \int d^nx \kappa \chi \phi ~,
\label{action1}
\end{equation}
where $\phi$ is a generic world-volume open string 
scalar field.
{}From our derivation, it is clear that $\chi$ field
with the effective action (\ref{action1}) gives the effective
description of the `long tube' IR closed strings at low 
energies.  The effective action (\ref{action1}) is 
identical to the one found in the noncommutative
field theory one-loop analysis \cite{seiberg1}.  

For the 
codimension one 24-brane, (\ref{edeg}) yields
\begin{equation}
{\cal A}_{\rm IR} = \left[ 
   \frac{\kappa^2\a'}{\sqrt{ p \circ p} }
 - \frac{\kappa^2\a'}{\sqrt{ p \circ p + 1/ \Lambda^2 } } \right] ~,  
\label{eff2}
\end{equation}
while for the codimension two 23-brane, we get
\begin{equation}
{\cal A}_{\rm IR} = \kappa^2 \left[ \log ( p \circ p) 
 - \log ( p \circ p + 1/ \Lambda^2  ) \right] ~.
\label{eff3}
\end{equation}
We note that (\ref{eff2}) and (\ref{eff3}) are the
same as the 1PI amplitudes found in Ref.~\cite{seiberg2}
for the extra low energy degrees of freedom.
From our derivation, it is clear that they also represent
the IR closed string degrees of freedom; they live in 
the bulk space-time while the open string degrees of 
freedom are confined on a codimension one and two D-brane,
respectively.  The extra dimensions found in 
Ref.~\cite{seiberg2} are indeed space-time dimensions
transversal to the brane, at least when the noncommutative
field theory under consideration derives from the decoupling
limit of open string theory.

%%%%%%%%%%%%%%%%%%%%%%%%%%%%%%%%%%%%%%%%%%%%%%%%%%%%%%%%%%%%%
%
\section{One-loop open superstring amplitudes}
%
%%%%%%%%%%%%%%%%%%%%%%%%%%%%%%%%%%%%%%%%%%%%%%%%%%%%%%%%%%%%%

In this section, we repeat the calculations of the 
previous section
for the case of open superstrings.  The main finding that
the space-time supersymmetry makes the two point amplitudes 
vanish regardless of the value of $B$ is consistent
with Ref.~\cite{susskind}.  In Ref.~\cite{susskind}, it was 
argued that for the supersymmetric gauge theories with
sixteen supercharges, the noncommutative IR singularities
do not show up.  In our present context with parallel
D-branes, we clearly have sixteen supercharges.  

The answer for the amplitude is given by
\be
\CA = \sum_{a=2}^4 (-1)^a
  \int_0^\infty \frac{dT}{T} (2\pi\a' T)^{-n/2} 
 \left\{\frac{f_a(q)}{f_1(q)}\right\}^8
T \int_0^T dy\langle V_1(p,y) V_2(-p,0) \rangle,
\ee
where the index $a$ labels spin structures. 
For the definitions of $f_a (q)$, see Ref.~\cite{pol}.
The vertex operators for massless gauge bosons are
\be
V(p,y) = \e_\m(\partial X^\m + i p\cdot\psi\psi^\m) e^{ipX}.
\ee
In addition to the terms in Eqs. \eq{pgaugeb} and \eq{npgaugeb}, 
we have the contraction of four world-sheet fermions.
After the summation over spin structures, however,
the two point function completely vanishes 
due to the Jacobi's fundamental formulae \cite{whw}. 
Naturally, one attributes this property to the space-time
supersymmetry.
In fact, only the terms with eight or more world-sheet fermions 
give nonzero contribution.

%%%%%%%%%%%%%%%%%%%%%%%%%%%%%%%%%%%%%%%%%%%%%%%%%%%%%%%%%%%%%
% 
\section{Discussions}
%
%%%%%%%%%%%%%%%%%%%%%%%%%%%%%%%%%%%%%%%%%%%%%%%%%%%%%%%%%%%%%

The term (\ref{winding}) should be present in the
world-sheet propagators as a consequence of the
boundary conditions (\ref{modbc}).  Its contribution
to nonplanar amplitudes is the crucial exponential factor
$\exp ( - (p \circ p) / 2 \pi \alpha^{\prime} T )$  
necessary for the emergence of IR closed string degrees 
of freedom.
Its existence, however, might seem rather 
puzzling from the open string theory point of view; the 
direction ${\rm Re } (z)$ is the spatial direction along which
the open string lies.  Therefore, it implies that there 
is a term in the mode expansion of $X$, which
is linearly proportional to ${\rm Re} (z)$.  For 
closed strings, this would usually
signal the presence of a non-trivial winding state, while
we are apparently considering open strings.  This behavior,
however, is consistent with Ref.~\cite{fischler} where 
the thermodynamic evidence for the `winding states' in 
noncommutative field theories is given.  As was shown 
from the calculations
in section 3, its contribution to amplitudes is from the 
corner of the open string moduli space where the dual closed 
string interpretation is appropriate.  From this point of view,
the `winding states' of Ref.~\cite{fischler} represent
nothing but closed string states.  We note that the authors
of Ref.~\cite{fischler} do not find any wrapping states,
which would correspond to higher extended objects than
strings.  

For the commutative field theories, the amplitude 
${\cal A}_{\rm IR}$ becomes a simple divergence involving 
the cutoff $\Lambda^2$ without the momentum dependence.  
In realistic models, via the Fischler-Susskind 
mechanism, the cutoff dependence 
in open string channel gets cancelled
by closed string sigma model divergence, ultimately
resulting the vanishing beta function.  In fact, the dual 
supergravity background geometries of 
commutative field theories have the asymptotic isometry
group isomorphic to the conformal group, a familiar
AdS/CFT correspondence \cite{adscft}; in this context,
we consider the perturbations around a conformal fixed point.
On the other hand, the conjectured dual supergravity backgrounds
of the noncommutative field theories do not have the asymptotic
isometry group isomorphic to the conformal group
\cite{aki,maldaruss}.  The emergence of the non-trivial
IR closed strings in noncommutative field theories
appears to be related to the nonconformality of the 
world-volume theory.  Naturally, the detailed investigation
of the Fischler-Susskind mechanism in the noncommutative
context along the line of Ref.~\cite{berenstein}
should reveal interesting physics.   

The noncommutative field theory calculations mimic the open
string calculations to a remarkable degree, as shown from the
analysis in this paper.  By turning the viewpoint around, 
one might consider using the noncommutative field theory as a
useful guide that provides us with a systematic organization
tool for the open/closed string loop diagrams.  In this spirit,
the disentangling of higher loop diagrams in noncommutative field 
theories via open string perturbation theory should be an
exciting venue, especially in relation to string field
theory \cite{witten,witten2}. 

\section*{Note added}

After the completion of the first version of this
paper, Ref.~\cite{bilal} appeared.  After reading 
\cite{bilal}, we found some calculational errors
in section 3.1 for the gauge boson amplitudes in the 
original
version of our paper.  The corrected
calculation revealed the $( \epsilon_1 \times p )
(\epsilon_2 \times p)$ part in Eq.~(\ref{npgaugeb}).
As noted in \cite{bilal}, some on-shell string 
calculations can be extended to off-shell in 
field theory limits.  Reinstalling the part that
vanishes in on-shell, the $ \left[ (\epsilon_1 \cdot
\epsilon_2 ) p^2 - (\epsilon_1 \cdot p )
( \epsilon_2 \cdot p) \right]$ part, in Eqs.~(\ref{pgaugeb})
and (\ref{npgaugeb}), we find that our gauge boson
amplitudes are identical to the ones given in
Ref.~\cite{bilal}.  
We emphasize that our boundary propagator 
expressions, Eqs.~(\ref{puha})-(\ref{puhaha}), are 
identical to Eqs.~(2.42)-(2.45) in
\cite{bilal} obtained by using the boundary state 
formalism.
To explicitly see this, we need variable changes
$2 \pi y = \log \left| \frac{\rho}{\rho^{\prime}} 
\right| $ and $2 \pi T = - \log k$. 

\section*{Acknowledgements}

We are grateful to Seungjoon Hyun, Kimyeong Lee, Hyeonjoon Shin 
and Piljin Yi for useful discussions. 
Y.K. would like to thank Chang-Yeong Lee for
useful correspondence, especially for bringing 
Ref.~\cite{fischler} to our attention.  While this paper was
being written, Ref.~\cite{overlap} came up, which has some
overlap with the material presented in Section 2 and 
Section 3.1.

\end{document}